\documentclass[aps,prl,twocolumn,superscriptaddress]{revtex4-1}

\usepackage{graphicx}

\usepackage{color}
\usepackage{braket}
\usepackage{hyperref}
\usepackage{amssymb}
\usepackage{float}

\usepackage{placeins}

\begin{document}

\title{Observation of superfluidity in a strongly correlated two-dimensional Fermi gas}

\author{Lennart Sobirey}
\email{lsobirey@physnet.uni-hamburg.de}
\affiliation{Institut f\"{u}r Laserphysik, Universit\"{a}t Hamburg}

\author{Niclas Luick}
\affiliation{Institut f\"{u}r Laserphysik, Universit\"{a}t Hamburg}
\affiliation{The Hamburg Centre for Ultrafast Imaging, Universit\"{a}t Hamburg, Luruper Chaussee 149, 22761 Hamburg}

\author{Markus Bohlen}
\affiliation{Institut f\"{u}r Laserphysik, Universit\"{a}t Hamburg}
\affiliation{The Hamburg Centre for Ultrafast Imaging, Universit\"{a}t Hamburg, Luruper Chaussee 149, 22761 Hamburg}
\affiliation{Laboratoire Kastler Brossel, ENS-Universit\'{e} PSL, CNRS, Sorbonne Universit\'{e}, Coll\`{e}ge de France, 24 rue Lhomond, 75005 Paris, France}

\author{Hauke Biss}
\affiliation{Institut f\"{u}r Laserphysik, Universit\"{a}t Hamburg}
\affiliation{The Hamburg Centre for Ultrafast Imaging, Universit\"{a}t Hamburg, Luruper Chaussee 149, 22761 Hamburg}

\author{Henning Moritz}
\affiliation{Institut f\"{u}r Laserphysik, Universit\"{a}t Hamburg}
\affiliation{The Hamburg Centre for Ultrafast Imaging, Universit\"{a}t Hamburg, Luruper Chaussee 149, 22761 Hamburg}

\author{Thomas Lompe}
\affiliation{Institut f\"{u}r Laserphysik, Universit\"{a}t Hamburg}

\begin{abstract}
Understanding how strongly correlated two-dimensional (2D) systems can give rise to unconventional superconductivity with high critical temperatures is one of the major unsolved problems in condensed matter physics. 
Ultracold 2D Fermi gases have emerged as clean and controllable model systems to study the interplay of strong correlations and reduced dimensionality, but direct evidence of superfluidity in these systems has been missing.
Here, we demonstrate superfluidity in an ultracold 2D Fermi gas by moving a periodic potential through the system and observing no dissipation below a critical velocity v$_{\rm c}$.
We measure v$_{\rm c}$ as a function of interaction strength and find a maximum in the crossover regime between bosonic and fermionic superfluidity.
Our measurement establishes ultracold Fermi gases as a powerful tool for studying the influence of reduced dimensionality on strongly correlated superfluids.
\end{abstract}

\maketitle

Reducing the dimensionality of a quantum system from three to two dimensions significantly modifies its physical properties.
One striking difference is the increased role of fluctuations in low-dimensional systems, which prevents long range phase coherence \cite{mermin1966absence} and makes 2D the marginal dimension for the existence of superfluidity \cite{kosterlitz1973ordering}. 
Hence it is quite surprising that in all known ambient-pressure high-T$_{\rm c}$ materials superconductivity occurs in two-dimensional structures, such as the copper-oxide layers in cuprates. 
Three decades after their discovery, the mechanism giving rise to superconductivity and the role of the reduced dimensionality in these systems are still under debate \cite{keimer2015quantum}.

Over the last years, ultracold 2D Fermi gases \cite{frohlich2011radio,sommer2012evolution,makhalov2014ground,ong2015spin,levinsen2015strongly,turlapov2017fermi} have been established as model systems that can provide insight into the interplay of strong correlations and reduced dimensionality \cite{murthy2018high,peppler2018quantum,holten2018anomalous,mitra2016phase,fenech2016thermodynamics,boettcher2016equation}.
Recent experiments have observed pair condensation \cite{ries2015observation} and phase coherence \cite{luick2019ideal} at low temperatures.
While these results suggest the presence of a superfluid, this has not been directly observed so far. 

In this work, we obtain definitive evidence for superfluidity in a 2D Fermi gas by observing frictionless flow below a critical velocity $v_{\rm c}$. 
We study the temperature-dependence of the critical velocity and observe the phase transition from the superfluid to the normal state at a critical temperature $T_{\rm c}$.
Finally, we measure the critical velocity as a function of interaction strength and show that the 2D Fermi gas is superfluid throughout the BEC-BCS crossover from deeply bound dimers to weakly bound Cooper pairs.

For our experiments, we use a Fermi gas of $N \approx 6000$ ultracold $^6$Li atoms in the lowest two hyperfine states, trapped in a box potential \cite{hueck2018two}.
The gas is tightly confined along the z-direction with a level spacing $\hbar \omega_{\rm z} \approx h\cdot 9.2\,\rm{kHz}$ that is larger than the thermal energy $k_{\rm B}T$ and the chemical potiential $\mu$ of the gas, which places our system in the quasi-2D regime \cite{Supp}.
\begin{figure}[H]
\begin{center}
\includegraphics[width=\linewidth]{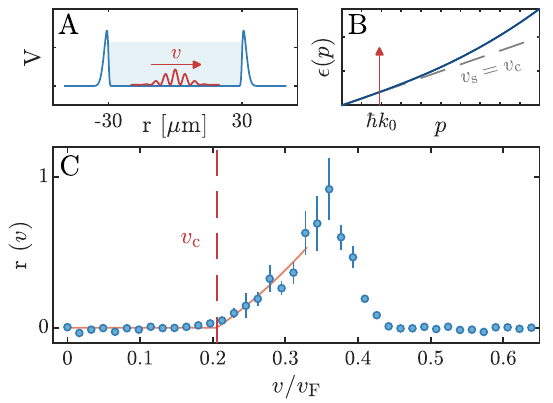}
\caption{\textbf{Measuring the critical velocity of a two-dimensional Fermi gas.}
(A) Sketch of the trapping potential. 
A 2D Fermi gas is trapped in a box potential (blue) projected through a high-resolution microscope objective. 
A periodic potential (red) can be moved through the gas at a variable velocity $v$.
(B) Sketch of the Bogoliubov dispersion of a superfluid Bose gas. 
At small momentum transfer $\hbar q$, excitations are phononic and the dispersion has a linear slope given by the speed of sound $v_{\rm s}$, while for higher momentum transfers single-particle excitations become dominant and the dispersion becomes quadratic. 
(C) Response $r(v)$ of a system at an interaction strength of ${\rm ln}(k_{\rm F}a_{\rm 2D})=-0.8$ to a moving optical lattice with wavevector $k_0 \approx 0.15\,k_{\rm F}$. 
While no dissipation occurs at low lattice velocities, there is a sharp increase in the response of the system above a critical velocity $v_{\rm c}$. 
Note that the moving lattice probes the dispersion relation of the system along a vertical line, as visualized by the red arrow in (B).
This results in a decrease of the response at high lattice velocities.
The critical velocity is extracted by a fit (red solid line) according to $r(v) = A\max(0, v^2-v_{\rm c}^2)$ \cite{astrakharchik2004motion}.
}
\label{Fig1}
\end{center}
\end{figure}
To show that the system is superfluid, we verify that it fulfills the Landau criterion \cite{landau1941theory,raman1999evidence,desbuquois2012superfluid,weimer2015critical,miller2007critical,ha2015roton}, which states that the dispersion $\epsilon(p)$ of a superfluid does not allow for the creation of excitations at velocities smaller than a nonzero critical velocity $v_{\rm c} = \min_p (\frac{\epsilon(p)}{p})$. 
Thus, an impurity moving through a superfluid with a velocity $v < v_{\rm c}$ creates no excitations, and the superfluid flows around it without friction.
We create such an impurity by interfering two red-detuned laser beams in the center of the trap, resulting in a sinusoidal potential whose wavelength can be tuned by adjusting the crossing angle of the two beams.
A frequency detuning $\Delta \nu$ between the two laser beams causes this optical lattice to move at a constant speed $v=L \Delta\nu$, where L is the spacing between two maxima of the periodic potential.

To measure the critical velocity in our system, we move the optical lattice through the gas at different velocities and observe the response of the system by measuring its momentum distribution $n(k)$. 
To obtain $n(k)$, we ramp the interaction strength to a value of ${\rm ln}(k_{\rm F}a_{\rm 2D}) = -2.8$, where $k_{\rm F} = \sqrt{4\pi n} = m v_{\rm F} / \hbar$ is the Fermi wavevector of a gas with density $n$ per spin state and $a_{\rm 2D}$ is the 2D scattering length \cite{petrov2001interatomic}.
At this interaction strength, the system is deep in the BEC regime, where the gas consists of weakly interacting dimers and it is straightforward to measure $n(k)$ using matter wave focusing \cite{murthy2014matter}. 
As the occupation of the lowest momentum modes decreases with increasing temperature, we define the response $r(v) = (n(k=0,v=0)/n(k=0,v))-1$ as a robust measure for the amount of energy that was imparted to the system by the moving potential \cite{Supp}.

A typical measurement of the response of the system as a function of lattice velocity is shown in Fig \ref{Fig1}\,C.  
We observe that as the velocity of the optical lattice is increased, the gas is unaffected until a critical velocity is reached and a sharp onset of dissipation occurs. 
In contrast to previous experiments \cite{raman1999evidence,miller2007critical,desbuquois2012superfluid,weimer2015critical}, we observe that the response decreases again at higher velocities. 
This is due to the fact that the optical lattice transfers a specific momentum $\hbar k_0 = \hbar 2\pi/L$ to the superfluid, whereas the impurity in Landau's gedankenexperiment can excite the system at all momenta.
Therefore, a moving optical lattice with varying velocity probes the dispersion relation of the gas on a vertical line of constant $p=\hbar k_0$.
This is visualized in Fig. \ref{Fig1}\,B.

\begin{figure}[htbp]
\begin{center}
\includegraphics[width=\linewidth]{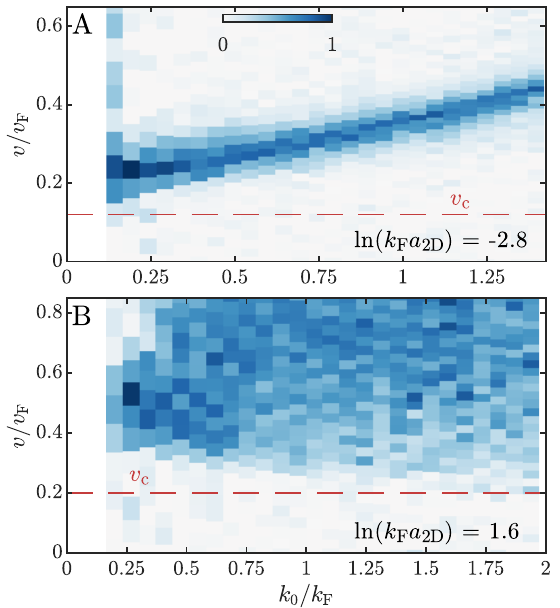}
\caption{\textbf{Phononic and pair breaking excitations in a 2D Fermi gas.}
(A) Response of a gas in the BEC regime to excitations with lattice wavevector $k_0$ and velocity $v$.
For small wavevectors, the moving lattice excites phononic modes at the sound velocity $v_{\rm s}$. 
For larger wavevectors, the peak in the heating rate moves to higher velocities as the dispersion deviates from the linear phononic branch and single-particle excitations become dominant.
(B) In the BCS regime, we observe a continuum of pair breaking excitations with a minimum of the onset velocity at $k_0 = 2\,k_{\rm F}$.
In both regimes, the heating rate is negligible for excitations that move slower than the critical velocity (red dashed lines, taken from Fig. \ref{Fig4}\,C).
To enhance the visibility of weaker excitations, each column has been linearly rescaled to range from 0 to 1.
}
\label{Fig2}
\end{center}
\end{figure}

We hence perform measurements at different spacings $L$ of the periodic potential, and thereby determine the response $r(v,k_0)$ as a function of both the lattice velocity $v$ and the lattice wavevector $k_0$.
In bosonic superfluids, the lowest velocity at which excitations can be created is found at small wavevectors. 
These long-wavelength excitations are phononic modes that are excited by an obstacle moving at a velocity close to the speed of sound of the system.
In BCS superfluids, phononic excitations at low $k_0$ can still be created, but the lowest onset velocity is found at $k_0\approx 2\,k_{\rm F}$.
This is due to pair breaking excitations, which can occur at all momenta but according to BCS theory can be excited with the lowest velocites at a wavevector of $2\,k_{\rm F}$.
Our measurements in the BEC (see Fig. \ref{Fig2}\,A) and BCS (see Fig. \ref{Fig2}\,B) regimes directly show this difference in the excitation spectra of bosonic and fermionic superfluids. 
For both interaction strengths, we clearly observe a critical velocity below which no excitations are created, which constitutes conclusive evidence of superfluidity.

\begin{figure}[htbp]
\begin{center}
\includegraphics[width=\linewidth]{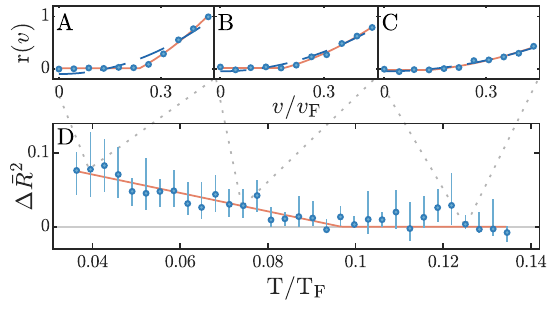}
\caption{\textbf{Observing the superfluid phase transition.}
(A,B,C) We determine the critical temperature of a gas in the BEC regime (${\rm ln}(k_{\rm F}a_{\rm 2D})=-2.9$) by measuring the response of the system to a moving periodic potential at different temperatures.
For a cold gas (A), the absence of dissipation at low velocities is followed by a sharp rise in the response of the gas at the critical velocity. 
As the temperature increases, the critical velocity is reduced (B) until dissipation occurs at arbitrarily small velocities once the temperature is above $T_{\rm c}$ (C).
The red solid lines are fits of the response $r(v)$ according to $r(v) = A\max(C, v^2-v_{\rm c}^2)$.
The blue dashed lines show fits where $v_{\rm c}$ is fixed to 0. 
(D) Difference $\Delta\bar{R}^2 = \bar{R}^2_{v_c>0} - \bar{R}^2_{v_c=0}$ of the adjusted coefficients of determination of the two fits.
For low temperature gases, allowing a finite critical velocity significantly improves the fit beyond the trivial effect of adding a free parameter and hence $\Delta\bar{R^2} > 0$. 
In contrast, this is not the case for high temperature systems above $T_{\rm c}$. 
We can therefore estimate $T_{\rm c}$ by extracting the temperature at which a nonzero $v_{\rm c}$ no longer improves the fit. 
For the measurement shown in this figure, we find that the phase transition to the normal state occurs at $T_{\rm c}/T_{\rm F} = 0.094 \pm 0.004_{\rm stat} \pm 0.02_{\rm sys}$ \cite{Supp}.
The errorbars show $1\sigma$ confidence intervals obtained by bootstrapping.}
\label{Fig3}
\end{center}
\end{figure}

Having established a measurement of the critical velocity, we now go on to determine the critical temperature $T_{\rm c}$ of a gas in the BEC regime (${\rm ln}(k_{\rm F}a_{\rm 2D})=-2.9$).
We achieve this by preparing gases at different temperatures \cite{Supp} and measuring the response of the system to the moving periodic potential.
With increasing temperature, we expect the phononic branch of the dispersion to broaden, and eventually become broad enough that excitations at arbitrarily small velocities can heat the gas.
This causes the critical velocity to decrease with temperature and vanish at $T=T_{\rm c}$.
Measurements of the response of the system for three different initial temperatures are shown in Fig. \ref{Fig3}\,A-C. 
While the sharp onset of dissipation at the critical velocity is clearly visible in the colder data, it disappears at higher temperatures, signalling the phase transition from a superfluid to a normal state. 
We extract a critical temperature of $T_{\rm c}/T_{\rm F} = 0.094 \pm 0.004_{\rm stat} \pm 0.02_{\rm sys}$ \cite{Supp}, which is in very good agreement with theoretical predictions \cite{levinsen2015strongly} and the observed onset of pair condensation \cite{ries2015observation}\footnote{Measurements performed at interaction strengths of up to ${\rm ln}(k_{\rm F}a_{\rm 2D}) = 1.1$ show the same qualitative behaviour \cite{Supp}, but we currently do not have a quantitatively accurate temperature determination for our homogenous system at these interaction strengths.}.

\begin{figure}[htbp]
\begin{center}
\includegraphics[width=\linewidth]{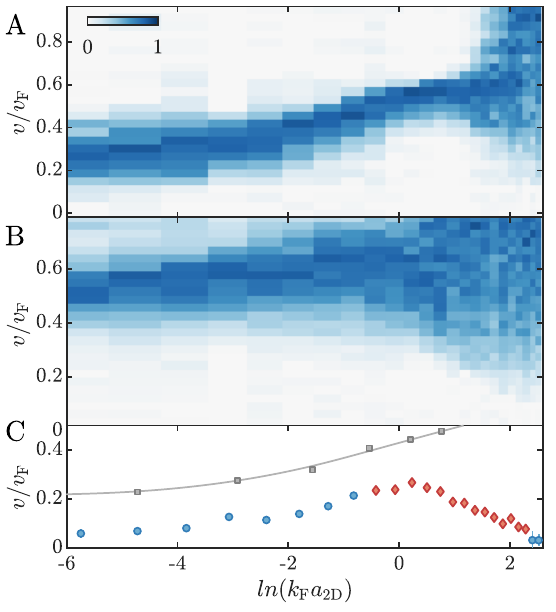}
\caption{\textbf{Interaction dependence of the critical velocity.}
(A,B) Response of a 2D Fermi gas to a moving lattice with wavevectors $k_0 \approx 0.3\,k_{\rm F}$ (A) and $k_0 \approx 2\,k_{\rm F}$ (B) at different interaction strengths.
In the BEC regime (${\rm ln}(k_{\rm F}a_{\rm 2D})<-1$), we observe a well defined excitation that corresponds to a sound mode for $k_0 \approx 0.3\,k_{\rm F}$ and single particle excitations for $k_0 \approx 2\,k_{\rm F}$.
When going to the BCS side of the crossover, the peak broadens into a continuum of pair breaking excitations.
To enhance the visibility of weaker excitations, each column has been linearly rescaled to range from 0 to 1.
(C) We determine the critical velocity as a function of interaction strength as the lower of the two onset velocities obtained from the data shown in (A,B).
In the BEC regime, the critical velocity is limited by excitations at small wavevectors (blue dots), while in the crossover the lowest onset velocities occur at $2\,k_{\rm F}$ (red diamonds).
We find that the 2D Fermi gas is superfluid throughout the 2D BEC-BCS crossover with the highest critical velocities found in the crossover regime at ${\rm ln}(k_{\rm F}a_{\rm 2D}) \approx 0$.
For comparison, we show the speed of sound $v_{\rm s}$ (grey squares) as measured in \cite{bohlen2020sound}, the grey line is a guide to the eye. 
The error bars denote the $1\sigma$ confidence intervals of the fit and are mostly smaller than the symbol size.}
\label{Fig4}
\end{center}
\end{figure}

In our final set of measurements, we study the evolution of the critical velocity in the crossover from a condensate of bosonic dimers to a BCS superfluid.
As shown in Fig. \ref{Fig2}, the lowest-lying excitations on the BEC side of the resonance are sound modes at small values of $k_0$, while for a BCS superfluid the minimum velocity for pair breaking occurs for excitations around $2\,k_{\rm F}$.
Hence, we measure the interaction dependence of the response $r(v)$ at two different lattice wavevectors of $k_0 \approx 0.3\,k_{\rm F}$ and $k_0 \approx 2\,k_{\rm F}$.
The results are shown in Fig. \ref{Fig4}\,A and B.

For a lattice wavevector of $k_0 \approx 0.3\,k_{\rm F}$, we clearly observe the presence of a well defined sound mode with an onset velocity that increases as a function of interaction strength.
In the crossover region (${\rm ln}(k_{\rm F}a_{\rm 2D}) \approx 0.5$), the peak smoothly broadens into a continuum as pair breaking becomes the dominant excitation in the system.
For $k_0 \approx 2\,k_{\rm F}$, the excitations on the BEC side are single particle excitations, with pair breaking taking over towards the BCS side of the resonance.
We fit the onset velocities for both data sets and use the smaller of the two values as the critical velocity of the system (see Fig. \ref{Fig4}\,C). 

The measured critical velocities scale with the speed of sound on the BEC side of the resonance, show a maximum in the crossover and decrease again as pair breaking becomes dominant in the BCS regime. 
The maximum of $v_{\rm c}$ at an interaction strength of ${\rm ln}(k_{\rm F}a_{\rm 2D}) \approx 0$ furthermore indicates that fermionic 2D superfluids are most stable in the strongly correlated crossover regime, in good agreement with the maximum of the critical temperature for pair condensation reported in \cite{ries2015observation}.

Our results establish 2D Fermi gases as ideal model systems to study how superfluidity is affected by the interplay of strong correlations and reduced dimensionality.
In particular, they can be used to study the transition from a superfluid to a strongly correlated pseudogap state above $T_{\rm c}$ in a much simpler and more accessible system than high-$T_{\rm c}$ superconductors.
Finally, the dimensionality of ultracold Fermi gases can be tuned continuously, making them uniquely suited to study the remarkable stability of the superfluid phase in the crossover from two to three dimensions.

\begin{acknowledgments}
We thank L. Mathey and V. Singh for valuable insights and stimulating discussions.
This work is supported by the European Union's Seventh Framework Programme (FP7/2007-2013) under grant agreement No. 335431 and by the Deutsche Forschungsgemeinschaft (DFG, German Research Foundation) in the framework of SFB 925 and the excellence cluster 'Advanced Imaging of Matter' - EXC 2056 - project ID 390715994.
M. Bohlen acknowledges support by Labex ICFP of \'{E}cole Normale Sup\'{e}rieure Paris.
\end{acknowledgments}

\bibliography{bib_vcrit}

\begin{thebibliography}{33}%
\makeatletter
\providecommand \@ifxundefined [1]{%
 \@ifx{#1\undefined}
}%
\providecommand \@ifnum [1]{%
 \ifnum #1\expandafter \@firstoftwo
 \else \expandafter \@secondoftwo
 \fi
}%
\providecommand \@ifx [1]{%
 \ifx #1\expandafter \@firstoftwo
 \else \expandafter \@secondoftwo
 \fi
}%
\providecommand \natexlab [1]{#1}%
\providecommand \enquote  [1]{``#1''}%
\providecommand \bibnamefont  [1]{#1}%
\providecommand \bibfnamefont [1]{#1}%
\providecommand \citenamefont [1]{#1}%
\providecommand \href@noop [0]{\@secondoftwo}%
\providecommand \href [0]{\begingroup \@sanitize@url \@href}%
\providecommand \@href[1]{\@@startlink{#1}\@@href}%
\providecommand \@@href[1]{\endgroup#1\@@endlink}%
\providecommand \@sanitize@url [0]{\catcode `\\12\catcode `\$12\catcode
  `\&12\catcode `\#12\catcode `\^12\catcode `\_12\catcode `\%12\relax}%
\providecommand \@@startlink[1]{}%
\providecommand \@@endlink[0]{}%
\providecommand \url  [0]{\begingroup\@sanitize@url \@url }%
\providecommand \@url [1]{\endgroup\@href {#1}{\urlprefix }}%
\providecommand \urlprefix  [0]{URL }%
\providecommand \Eprint [0]{\href }%
\providecommand \doibase [0]{http://dx.doi.org/}%
\providecommand \selectlanguage [0]{\@gobble}%
\providecommand \bibinfo  [0]{\@secondoftwo}%
\providecommand \bibfield  [0]{\@secondoftwo}%
\providecommand \translation [1]{[#1]}%
\providecommand \BibitemOpen [0]{}%
\providecommand \bibitemStop [0]{}%
\providecommand \bibitemNoStop [0]{.\EOS\space}%
\providecommand \EOS [0]{\spacefactor3000\relax}%
\providecommand \BibitemShut  [1]{\csname bibitem#1\endcsname}%
\let\auto@bib@innerbib\@empty
\bibitem [{\citenamefont {Mermin}\ and\ \citenamefont
  {Wagner}(1966)}]{mermin1966absence}%
  \BibitemOpen
  \bibfield  {author} {\bibinfo {author} {\bibfnamefont {N.~D.}\ \bibnamefont
  {Mermin}}\ and\ \bibinfo {author} {\bibfnamefont {H.}~\bibnamefont
  {Wagner}},\ }\href@noop {} {\bibfield  {journal} {\bibinfo  {journal}
  {Physical Review Letters}\ }\textbf {\bibinfo {volume} {17}},\ \bibinfo
  {pages} {1133} (\bibinfo {year} {1966})}\BibitemShut {NoStop}%
\bibitem [{\citenamefont {Kosterlitz}\ and\ \citenamefont
  {Thouless}(1973)}]{kosterlitz1973ordering}%
  \BibitemOpen
  \bibfield  {author} {\bibinfo {author} {\bibfnamefont {J.~M.}\ \bibnamefont
  {Kosterlitz}}\ and\ \bibinfo {author} {\bibfnamefont {D.~J.}\ \bibnamefont
  {Thouless}},\ }\href@noop {} {\bibfield  {journal} {\bibinfo  {journal}
  {Journal of Physics C: Solid State Physics}\ }\textbf {\bibinfo {volume}
  {6}},\ \bibinfo {pages} {1181} (\bibinfo {year} {1973})}\BibitemShut
  {NoStop}%
\bibitem [{\citenamefont {Keimer}\ \emph {et~al.}(2015)\citenamefont {Keimer},
  \citenamefont {Kivelson}, \citenamefont {Norman}, \citenamefont {Uchida},\
  and\ \citenamefont {Zaanen}}]{keimer2015quantum}%
  \BibitemOpen
  \bibfield  {author} {\bibinfo {author} {\bibfnamefont {B.}~\bibnamefont
  {Keimer}}, \bibinfo {author} {\bibfnamefont {S.~A.}\ \bibnamefont
  {Kivelson}}, \bibinfo {author} {\bibfnamefont {M.~R.}\ \bibnamefont
  {Norman}}, \bibinfo {author} {\bibfnamefont {S.}~\bibnamefont {Uchida}}, \
  and\ \bibinfo {author} {\bibfnamefont {J.}~\bibnamefont {Zaanen}},\
  }\href@noop {} {\bibfield  {journal} {\bibinfo  {journal} {Nature}\ }\textbf
  {\bibinfo {volume} {518}},\ \bibinfo {pages} {179} (\bibinfo {year}
  {2015})}\BibitemShut {NoStop}%
\bibitem [{\citenamefont {Fr{\"o}hlich}\ \emph {et~al.}(2011)\citenamefont
  {Fr{\"o}hlich}, \citenamefont {Feld}, \citenamefont {Vogt}, \citenamefont
  {Koschorreck}, \citenamefont {Zwerger},\ and\ \citenamefont
  {K{\"o}hl}}]{frohlich2011radio}%
  \BibitemOpen
  \bibfield  {author} {\bibinfo {author} {\bibfnamefont {B.}~\bibnamefont
  {Fr{\"o}hlich}}, \bibinfo {author} {\bibfnamefont {M.}~\bibnamefont {Feld}},
  \bibinfo {author} {\bibfnamefont {E.}~\bibnamefont {Vogt}}, \bibinfo {author}
  {\bibfnamefont {M.}~\bibnamefont {Koschorreck}}, \bibinfo {author}
  {\bibfnamefont {W.}~\bibnamefont {Zwerger}}, \ and\ \bibinfo {author}
  {\bibfnamefont {M.}~\bibnamefont {K{\"o}hl}},\ }\href@noop {} {\bibfield
  {journal} {\bibinfo  {journal} {Physical review letters}\ }\textbf {\bibinfo
  {volume} {106}},\ \bibinfo {pages} {105301} (\bibinfo {year}
  {2011})}\BibitemShut {NoStop}%
\bibitem [{\citenamefont {Sommer}\ \emph {et~al.}(2012)\citenamefont {Sommer},
  \citenamefont {Cheuk}, \citenamefont {Ku}, \citenamefont {Bakr},\ and\
  \citenamefont {Zwierlein}}]{sommer2012evolution}%
  \BibitemOpen
  \bibfield  {author} {\bibinfo {author} {\bibfnamefont {A.~T.}\ \bibnamefont
  {Sommer}}, \bibinfo {author} {\bibfnamefont {L.~W.}\ \bibnamefont {Cheuk}},
  \bibinfo {author} {\bibfnamefont {M.~J.}\ \bibnamefont {Ku}}, \bibinfo
  {author} {\bibfnamefont {W.~S.}\ \bibnamefont {Bakr}}, \ and\ \bibinfo
  {author} {\bibfnamefont {M.~W.}\ \bibnamefont {Zwierlein}},\ }\href@noop {}
  {\bibfield  {journal} {\bibinfo  {journal} {Physical review letters}\
  }\textbf {\bibinfo {volume} {108}},\ \bibinfo {pages} {045302} (\bibinfo
  {year} {2012})}\BibitemShut {NoStop}%
\bibitem [{\citenamefont {Makhalov}\ \emph {et~al.}(2014)\citenamefont
  {Makhalov}, \citenamefont {Martiyanov},\ and\ \citenamefont
  {Turlapov}}]{makhalov2014ground}%
  \BibitemOpen
  \bibfield  {author} {\bibinfo {author} {\bibfnamefont {V.}~\bibnamefont
  {Makhalov}}, \bibinfo {author} {\bibfnamefont {K.}~\bibnamefont
  {Martiyanov}}, \ and\ \bibinfo {author} {\bibfnamefont {A.}~\bibnamefont
  {Turlapov}},\ }\href@noop {} {\bibfield  {journal} {\bibinfo  {journal}
  {Physical review letters}\ }\textbf {\bibinfo {volume} {112}},\ \bibinfo
  {pages} {045301} (\bibinfo {year} {2014})}\BibitemShut {NoStop}%
\bibitem [{\citenamefont {Ong}\ \emph {et~al.}(2015)\citenamefont {Ong},
  \citenamefont {Cheng}, \citenamefont {Arakelyan},\ and\ \citenamefont
  {Thomas}}]{ong2015spin}%
  \BibitemOpen
  \bibfield  {author} {\bibinfo {author} {\bibfnamefont {W.}~\bibnamefont
  {Ong}}, \bibinfo {author} {\bibfnamefont {C.}~\bibnamefont {Cheng}}, \bibinfo
  {author} {\bibfnamefont {I.}~\bibnamefont {Arakelyan}}, \ and\ \bibinfo
  {author} {\bibfnamefont {J.}~\bibnamefont {Thomas}},\ }\href@noop {}
  {\bibfield  {journal} {\bibinfo  {journal} {Physical review letters}\
  }\textbf {\bibinfo {volume} {114}},\ \bibinfo {pages} {110403} (\bibinfo
  {year} {2015})}\BibitemShut {NoStop}%
\bibitem [{\citenamefont {Levinsen}\ and\ \citenamefont
  {Parish}(2015)}]{levinsen2015strongly}%
  \BibitemOpen
  \bibfield  {author} {\bibinfo {author} {\bibfnamefont {J.}~\bibnamefont
  {Levinsen}}\ and\ \bibinfo {author} {\bibfnamefont {M.~M.}\ \bibnamefont
  {Parish}},\ }in\ \href@noop {} {\emph {\bibinfo {booktitle} {Annual review of
  cold atoms and molecules}}}\ (\bibinfo  {publisher} {World Scientific},\
  \bibinfo {year} {2015})\ pp.\ \bibinfo {pages} {1--75}\BibitemShut {NoStop}%
\bibitem [{\citenamefont {Turlapov}\ and\ \citenamefont
  {Kagan}(2017)}]{turlapov2017fermi}%
  \BibitemOpen
  \bibfield  {author} {\bibinfo {author} {\bibfnamefont {A.}~\bibnamefont
  {Turlapov}}\ and\ \bibinfo {author} {\bibfnamefont {M.~Y.}\ \bibnamefont
  {Kagan}},\ }\href@noop {} {\bibfield  {journal} {\bibinfo  {journal} {Journal
  of Physics: Condensed Matter}\ }\textbf {\bibinfo {volume} {29}},\ \bibinfo
  {pages} {383004} (\bibinfo {year} {2017})}\BibitemShut {NoStop}%
\bibitem [{\citenamefont {Murthy}\ \emph {et~al.}(2018)\citenamefont {Murthy},
  \citenamefont {Neidig}, \citenamefont {Klemt}, \citenamefont {Bayha},
  \citenamefont {Boettcher}, \citenamefont {Enss}, \citenamefont {Holten},
  \citenamefont {Z{\"u}rn}, \citenamefont {Preiss},\ and\ \citenamefont
  {Jochim}}]{murthy2018high}%
  \BibitemOpen
  \bibfield  {author} {\bibinfo {author} {\bibfnamefont {P.~A.}\ \bibnamefont
  {Murthy}}, \bibinfo {author} {\bibfnamefont {M.}~\bibnamefont {Neidig}},
  \bibinfo {author} {\bibfnamefont {R.}~\bibnamefont {Klemt}}, \bibinfo
  {author} {\bibfnamefont {L.}~\bibnamefont {Bayha}}, \bibinfo {author}
  {\bibfnamefont {I.}~\bibnamefont {Boettcher}}, \bibinfo {author}
  {\bibfnamefont {T.}~\bibnamefont {Enss}}, \bibinfo {author} {\bibfnamefont
  {M.}~\bibnamefont {Holten}}, \bibinfo {author} {\bibfnamefont
  {G.}~\bibnamefont {Z{\"u}rn}}, \bibinfo {author} {\bibfnamefont {P.~M.}\
  \bibnamefont {Preiss}}, \ and\ \bibinfo {author} {\bibfnamefont
  {S.}~\bibnamefont {Jochim}},\ }\href@noop {} {\bibfield  {journal} {\bibinfo
  {journal} {Science}\ }\textbf {\bibinfo {volume} {359}},\ \bibinfo {pages}
  {452} (\bibinfo {year} {2018})}\BibitemShut {NoStop}%
\bibitem [{\citenamefont {Peppler}\ \emph {et~al.}(2018)\citenamefont
  {Peppler}, \citenamefont {Dyke}, \citenamefont {Zamorano}, \citenamefont
  {Herrera}, \citenamefont {Hoinka},\ and\ \citenamefont
  {Vale}}]{peppler2018quantum}%
  \BibitemOpen
  \bibfield  {author} {\bibinfo {author} {\bibfnamefont {T.}~\bibnamefont
  {Peppler}}, \bibinfo {author} {\bibfnamefont {P.}~\bibnamefont {Dyke}},
  \bibinfo {author} {\bibfnamefont {M.}~\bibnamefont {Zamorano}}, \bibinfo
  {author} {\bibfnamefont {I.}~\bibnamefont {Herrera}}, \bibinfo {author}
  {\bibfnamefont {S.}~\bibnamefont {Hoinka}}, \ and\ \bibinfo {author}
  {\bibfnamefont {C.}~\bibnamefont {Vale}},\ }\href@noop {} {\bibfield
  {journal} {\bibinfo  {journal} {Physical review letters}\ }\textbf {\bibinfo
  {volume} {121}},\ \bibinfo {pages} {120402} (\bibinfo {year}
  {2018})}\BibitemShut {NoStop}%
\bibitem [{\citenamefont {Holten}\ \emph {et~al.}(2018)\citenamefont {Holten},
  \citenamefont {Bayha}, \citenamefont {Klein}, \citenamefont {Murthy},
  \citenamefont {Preiss},\ and\ \citenamefont {Jochim}}]{holten2018anomalous}%
  \BibitemOpen
  \bibfield  {author} {\bibinfo {author} {\bibfnamefont {M.}~\bibnamefont
  {Holten}}, \bibinfo {author} {\bibfnamefont {L.}~\bibnamefont {Bayha}},
  \bibinfo {author} {\bibfnamefont {A.~C.}\ \bibnamefont {Klein}}, \bibinfo
  {author} {\bibfnamefont {P.~A.}\ \bibnamefont {Murthy}}, \bibinfo {author}
  {\bibfnamefont {P.~M.}\ \bibnamefont {Preiss}}, \ and\ \bibinfo {author}
  {\bibfnamefont {S.}~\bibnamefont {Jochim}},\ }\href@noop {} {\bibfield
  {journal} {\bibinfo  {journal} {Physical review letters}\ }\textbf {\bibinfo
  {volume} {121}},\ \bibinfo {pages} {120401} (\bibinfo {year}
  {2018})}\BibitemShut {NoStop}%
\bibitem [{\citenamefont {Mitra}\ \emph {et~al.}(2016)\citenamefont {Mitra},
  \citenamefont {Brown}, \citenamefont {Schau{\ss}}, \citenamefont {Kondov},\
  and\ \citenamefont {Bakr}}]{mitra2016phase}%
  \BibitemOpen
  \bibfield  {author} {\bibinfo {author} {\bibfnamefont {D.}~\bibnamefont
  {Mitra}}, \bibinfo {author} {\bibfnamefont {P.~T.}\ \bibnamefont {Brown}},
  \bibinfo {author} {\bibfnamefont {P.}~\bibnamefont {Schau{\ss}}}, \bibinfo
  {author} {\bibfnamefont {S.~S.}\ \bibnamefont {Kondov}}, \ and\ \bibinfo
  {author} {\bibfnamefont {W.~S.}\ \bibnamefont {Bakr}},\ }\href@noop {}
  {\bibfield  {journal} {\bibinfo  {journal} {Physical review letters}\
  }\textbf {\bibinfo {volume} {117}},\ \bibinfo {pages} {093601} (\bibinfo
  {year} {2016})}\BibitemShut {NoStop}%
\bibitem [{\citenamefont {Fenech}\ \emph {et~al.}(2016)\citenamefont {Fenech},
  \citenamefont {Dyke}, \citenamefont {Peppler}, \citenamefont {Lingham},
  \citenamefont {Hoinka}, \citenamefont {Hu},\ and\ \citenamefont
  {Vale}}]{fenech2016thermodynamics}%
  \BibitemOpen
  \bibfield  {author} {\bibinfo {author} {\bibfnamefont {K.}~\bibnamefont
  {Fenech}}, \bibinfo {author} {\bibfnamefont {P.}~\bibnamefont {Dyke}},
  \bibinfo {author} {\bibfnamefont {T.}~\bibnamefont {Peppler}}, \bibinfo
  {author} {\bibfnamefont {M.}~\bibnamefont {Lingham}}, \bibinfo {author}
  {\bibfnamefont {S.}~\bibnamefont {Hoinka}}, \bibinfo {author} {\bibfnamefont
  {H.}~\bibnamefont {Hu}}, \ and\ \bibinfo {author} {\bibfnamefont
  {C.}~\bibnamefont {Vale}},\ }\href@noop {} {\bibfield  {journal} {\bibinfo
  {journal} {Physical review letters}\ }\textbf {\bibinfo {volume} {116}},\
  \bibinfo {pages} {045302} (\bibinfo {year} {2016})}\BibitemShut {NoStop}%
\bibitem [{\citenamefont {Boettcher}\ \emph {et~al.}(2016)\citenamefont
  {Boettcher}, \citenamefont {Bayha}, \citenamefont {Kedar}, \citenamefont
  {Murthy}, \citenamefont {Neidig}, \citenamefont {Ries}, \citenamefont {Wenz},
  \citenamefont {Zuern}, \citenamefont {Jochim},\ and\ \citenamefont
  {Enss}}]{boettcher2016equation}%
  \BibitemOpen
  \bibfield  {author} {\bibinfo {author} {\bibfnamefont {I.}~\bibnamefont
  {Boettcher}}, \bibinfo {author} {\bibfnamefont {L.}~\bibnamefont {Bayha}},
  \bibinfo {author} {\bibfnamefont {D.}~\bibnamefont {Kedar}}, \bibinfo
  {author} {\bibfnamefont {P.}~\bibnamefont {Murthy}}, \bibinfo {author}
  {\bibfnamefont {M.}~\bibnamefont {Neidig}}, \bibinfo {author} {\bibfnamefont
  {M.}~\bibnamefont {Ries}}, \bibinfo {author} {\bibfnamefont {A.}~\bibnamefont
  {Wenz}}, \bibinfo {author} {\bibfnamefont {G.}~\bibnamefont {Zuern}},
  \bibinfo {author} {\bibfnamefont {S.}~\bibnamefont {Jochim}}, \ and\ \bibinfo
  {author} {\bibfnamefont {T.}~\bibnamefont {Enss}},\ }\href@noop {} {\bibfield
   {journal} {\bibinfo  {journal} {Physical review letters}\ }\textbf {\bibinfo
  {volume} {116}},\ \bibinfo {pages} {045303} (\bibinfo {year}
  {2016})}\BibitemShut {NoStop}%
\bibitem [{\citenamefont {Ries}\ \emph {et~al.}(2015)\citenamefont {Ries},
  \citenamefont {Wenz}, \citenamefont {Z{\"u}rn}, \citenamefont {Bayha},
  \citenamefont {Boettcher}, \citenamefont {Kedar}, \citenamefont {Murthy},
  \citenamefont {Neidig}, \citenamefont {Lompe},\ and\ \citenamefont
  {Jochim}}]{ries2015observation}%
  \BibitemOpen
  \bibfield  {author} {\bibinfo {author} {\bibfnamefont {M.}~\bibnamefont
  {Ries}}, \bibinfo {author} {\bibfnamefont {A.}~\bibnamefont {Wenz}}, \bibinfo
  {author} {\bibfnamefont {G.}~\bibnamefont {Z{\"u}rn}}, \bibinfo {author}
  {\bibfnamefont {L.}~\bibnamefont {Bayha}}, \bibinfo {author} {\bibfnamefont
  {I.}~\bibnamefont {Boettcher}}, \bibinfo {author} {\bibfnamefont
  {D.}~\bibnamefont {Kedar}}, \bibinfo {author} {\bibfnamefont
  {P.}~\bibnamefont {Murthy}}, \bibinfo {author} {\bibfnamefont
  {M.}~\bibnamefont {Neidig}}, \bibinfo {author} {\bibfnamefont
  {T.}~\bibnamefont {Lompe}}, \ and\ \bibinfo {author} {\bibfnamefont
  {S.}~\bibnamefont {Jochim}},\ }\href@noop {} {\bibfield  {journal} {\bibinfo
  {journal} {Physical review letters}\ }\textbf {\bibinfo {volume} {114}},\
  \bibinfo {pages} {230401} (\bibinfo {year} {2015})}\BibitemShut {NoStop}%
\bibitem [{\citenamefont {Luick}\ \emph {et~al.}(2019)\citenamefont {Luick},
  \citenamefont {Sobirey}, \citenamefont {Bohlen}, \citenamefont {Singh},
  \citenamefont {Mathey}, \citenamefont {Lompe},\ and\ \citenamefont
  {Moritz}}]{luick2019ideal}%
  \BibitemOpen
  \bibfield  {author} {\bibinfo {author} {\bibfnamefont {N.}~\bibnamefont
  {Luick}}, \bibinfo {author} {\bibfnamefont {L.}~\bibnamefont {Sobirey}},
  \bibinfo {author} {\bibfnamefont {M.}~\bibnamefont {Bohlen}}, \bibinfo
  {author} {\bibfnamefont {V.~P.}\ \bibnamefont {Singh}}, \bibinfo {author}
  {\bibfnamefont {L.}~\bibnamefont {Mathey}}, \bibinfo {author} {\bibfnamefont
  {T.}~\bibnamefont {Lompe}}, \ and\ \bibinfo {author} {\bibfnamefont
  {H.}~\bibnamefont {Moritz}},\ }\href@noop {} {\bibfield  {journal} {\bibinfo
  {journal} {arXiv preprint arXiv:1908.09776}\ } (\bibinfo {year}
  {2019})}\BibitemShut {NoStop}%
\bibitem [{\citenamefont {Hueck}\ \emph {et~al.}(2018)\citenamefont {Hueck},
  \citenamefont {Luick}, \citenamefont {Sobirey}, \citenamefont {Siegl},
  \citenamefont {Lompe},\ and\ \citenamefont {Moritz}}]{hueck2018two}%
  \BibitemOpen
  \bibfield  {author} {\bibinfo {author} {\bibfnamefont {K.}~\bibnamefont
  {Hueck}}, \bibinfo {author} {\bibfnamefont {N.}~\bibnamefont {Luick}},
  \bibinfo {author} {\bibfnamefont {L.}~\bibnamefont {Sobirey}}, \bibinfo
  {author} {\bibfnamefont {J.}~\bibnamefont {Siegl}}, \bibinfo {author}
  {\bibfnamefont {T.}~\bibnamefont {Lompe}}, \ and\ \bibinfo {author}
  {\bibfnamefont {H.}~\bibnamefont {Moritz}},\ }\href@noop {} {\bibfield
  {journal} {\bibinfo  {journal} {Physical review letters}\ }\textbf {\bibinfo
  {volume} {120}},\ \bibinfo {pages} {060402} (\bibinfo {year}
  {2018})}\BibitemShut {NoStop}%
\bibitem [{Sup()}]{Supp}%
  \BibitemOpen
  \href@noop {} {}\bibinfo {note} {See Supplementary Materials}\BibitemShut
  {NoStop}%
\bibitem [{\citenamefont {Astrakharchik}\ and\ \citenamefont
  {Pitaevskii}(2004)}]{astrakharchik2004motion}%
  \BibitemOpen
  \bibfield  {author} {\bibinfo {author} {\bibfnamefont {G.}~\bibnamefont
  {Astrakharchik}}\ and\ \bibinfo {author} {\bibfnamefont {L.}~\bibnamefont
  {Pitaevskii}},\ }\href@noop {} {\bibfield  {journal} {\bibinfo  {journal}
  {Physical Review A}\ }\textbf {\bibinfo {volume} {70}},\ \bibinfo {pages}
  {013608} (\bibinfo {year} {2004})}\BibitemShut {NoStop}%
\bibitem [{\citenamefont {Landau}(1941)}]{landau1941theory}%
  \BibitemOpen
  \bibfield  {author} {\bibinfo {author} {\bibfnamefont {L.}~\bibnamefont
  {Landau}},\ }\href@noop {} {\bibfield  {journal} {\bibinfo  {journal}
  {Physical Review}\ }\textbf {\bibinfo {volume} {60}},\ \bibinfo {pages} {356}
  (\bibinfo {year} {1941})}\BibitemShut {NoStop}%
\bibitem [{\citenamefont {Raman}\ \emph {et~al.}(1999)\citenamefont {Raman},
  \citenamefont {K{\"o}hl}, \citenamefont {Onofrio}, \citenamefont {Durfee},
  \citenamefont {Kuklewicz}, \citenamefont {Hadzibabic},\ and\ \citenamefont
  {Ketterle}}]{raman1999evidence}%
  \BibitemOpen
  \bibfield  {author} {\bibinfo {author} {\bibfnamefont {C.}~\bibnamefont
  {Raman}}, \bibinfo {author} {\bibfnamefont {M.}~\bibnamefont {K{\"o}hl}},
  \bibinfo {author} {\bibfnamefont {R.}~\bibnamefont {Onofrio}}, \bibinfo
  {author} {\bibfnamefont {D.}~\bibnamefont {Durfee}}, \bibinfo {author}
  {\bibfnamefont {C.}~\bibnamefont {Kuklewicz}}, \bibinfo {author}
  {\bibfnamefont {Z.}~\bibnamefont {Hadzibabic}}, \ and\ \bibinfo {author}
  {\bibfnamefont {W.}~\bibnamefont {Ketterle}},\ }\href@noop {} {\bibfield
  {journal} {\bibinfo  {journal} {Physical Review Letters}\ }\textbf {\bibinfo
  {volume} {83}},\ \bibinfo {pages} {2502} (\bibinfo {year}
  {1999})}\BibitemShut {NoStop}%
\bibitem [{\citenamefont {Desbuquois}\ \emph {et~al.}(2012)\citenamefont
  {Desbuquois}, \citenamefont {Chomaz}, \citenamefont {Yefsah}, \citenamefont
  {L{\'e}onard}, \citenamefont {Beugnon}, \citenamefont {Weitenberg},\ and\
  \citenamefont {Dalibard}}]{desbuquois2012superfluid}%
  \BibitemOpen
  \bibfield  {author} {\bibinfo {author} {\bibfnamefont {R.}~\bibnamefont
  {Desbuquois}}, \bibinfo {author} {\bibfnamefont {L.}~\bibnamefont {Chomaz}},
  \bibinfo {author} {\bibfnamefont {T.}~\bibnamefont {Yefsah}}, \bibinfo
  {author} {\bibfnamefont {J.}~\bibnamefont {L{\'e}onard}}, \bibinfo {author}
  {\bibfnamefont {J.}~\bibnamefont {Beugnon}}, \bibinfo {author} {\bibfnamefont
  {C.}~\bibnamefont {Weitenberg}}, \ and\ \bibinfo {author} {\bibfnamefont
  {J.}~\bibnamefont {Dalibard}},\ }\href@noop {} {\bibfield  {journal}
  {\bibinfo  {journal} {Nature Physics}\ }\textbf {\bibinfo {volume} {8}},\
  \bibinfo {pages} {645} (\bibinfo {year} {2012})}\BibitemShut {NoStop}%
\bibitem [{\citenamefont {Weimer}\ \emph {et~al.}(2015)\citenamefont {Weimer},
  \citenamefont {Morgener}, \citenamefont {Singh}, \citenamefont {Siegl},
  \citenamefont {Hueck}, \citenamefont {Luick}, \citenamefont {Mathey},\ and\
  \citenamefont {Moritz}}]{weimer2015critical}%
  \BibitemOpen
  \bibfield  {author} {\bibinfo {author} {\bibfnamefont {W.}~\bibnamefont
  {Weimer}}, \bibinfo {author} {\bibfnamefont {K.}~\bibnamefont {Morgener}},
  \bibinfo {author} {\bibfnamefont {V.~P.}\ \bibnamefont {Singh}}, \bibinfo
  {author} {\bibfnamefont {J.}~\bibnamefont {Siegl}}, \bibinfo {author}
  {\bibfnamefont {K.}~\bibnamefont {Hueck}}, \bibinfo {author} {\bibfnamefont
  {N.}~\bibnamefont {Luick}}, \bibinfo {author} {\bibfnamefont
  {L.}~\bibnamefont {Mathey}}, \ and\ \bibinfo {author} {\bibfnamefont
  {H.}~\bibnamefont {Moritz}},\ }\href@noop {} {\bibfield  {journal} {\bibinfo
  {journal} {Physical review letters}\ }\textbf {\bibinfo {volume} {114}},\
  \bibinfo {pages} {095301} (\bibinfo {year} {2015})}\BibitemShut {NoStop}%
\bibitem [{\citenamefont {Miller}\ \emph {et~al.}(2007)\citenamefont {Miller},
  \citenamefont {Chin}, \citenamefont {Stan}, \citenamefont {Liu},
  \citenamefont {Setiawan}, \citenamefont {Sanner},\ and\ \citenamefont
  {Ketterle}}]{miller2007critical}%
  \BibitemOpen
  \bibfield  {author} {\bibinfo {author} {\bibfnamefont {D.}~\bibnamefont
  {Miller}}, \bibinfo {author} {\bibfnamefont {J.}~\bibnamefont {Chin}},
  \bibinfo {author} {\bibfnamefont {C.}~\bibnamefont {Stan}}, \bibinfo {author}
  {\bibfnamefont {Y.}~\bibnamefont {Liu}}, \bibinfo {author} {\bibfnamefont
  {W.}~\bibnamefont {Setiawan}}, \bibinfo {author} {\bibfnamefont
  {C.}~\bibnamefont {Sanner}}, \ and\ \bibinfo {author} {\bibfnamefont
  {W.}~\bibnamefont {Ketterle}},\ }\href@noop {} {\bibfield  {journal}
  {\bibinfo  {journal} {Physical review letters}\ }\textbf {\bibinfo {volume}
  {99}},\ \bibinfo {pages} {070402} (\bibinfo {year} {2007})}\BibitemShut
  {NoStop}%
\bibitem [{\citenamefont {Ha}\ \emph {et~al.}(2015)\citenamefont {Ha},
  \citenamefont {Clark}, \citenamefont {Parker}, \citenamefont {Anderson},\
  and\ \citenamefont {Chin}}]{ha2015roton}%
  \BibitemOpen
  \bibfield  {author} {\bibinfo {author} {\bibfnamefont {L.-C.}\ \bibnamefont
  {Ha}}, \bibinfo {author} {\bibfnamefont {L.~W.}\ \bibnamefont {Clark}},
  \bibinfo {author} {\bibfnamefont {C.~V.}\ \bibnamefont {Parker}}, \bibinfo
  {author} {\bibfnamefont {B.~M.}\ \bibnamefont {Anderson}}, \ and\ \bibinfo
  {author} {\bibfnamefont {C.}~\bibnamefont {Chin}},\ }\href@noop {} {\bibfield
   {journal} {\bibinfo  {journal} {Physical review letters}\ }\textbf {\bibinfo
  {volume} {114}},\ \bibinfo {pages} {055301} (\bibinfo {year}
  {2015})}\BibitemShut {NoStop}%
\bibitem [{\citenamefont {Petrov}\ and\ \citenamefont
  {Shlyapnikov}(2001)}]{petrov2001interatomic}%
  \BibitemOpen
  \bibfield  {author} {\bibinfo {author} {\bibfnamefont {D.}~\bibnamefont
  {Petrov}}\ and\ \bibinfo {author} {\bibfnamefont {G.}~\bibnamefont
  {Shlyapnikov}},\ }\href@noop {} {\bibfield  {journal} {\bibinfo  {journal}
  {Physical Review A}\ }\textbf {\bibinfo {volume} {64}},\ \bibinfo {pages}
  {012706} (\bibinfo {year} {2001})}\BibitemShut {NoStop}%
\bibitem [{\citenamefont {Murthy}\ \emph {et~al.}(2014)\citenamefont {Murthy},
  \citenamefont {Kedar}, \citenamefont {Lompe}, \citenamefont {Neidig},
  \citenamefont {Ries}, \citenamefont {Wenz}, \citenamefont {Z{\"u}rn},\ and\
  \citenamefont {Jochim}}]{murthy2014matter}%
  \BibitemOpen
  \bibfield  {author} {\bibinfo {author} {\bibfnamefont {P.}~\bibnamefont
  {Murthy}}, \bibinfo {author} {\bibfnamefont {D.}~\bibnamefont {Kedar}},
  \bibinfo {author} {\bibfnamefont {T.}~\bibnamefont {Lompe}}, \bibinfo
  {author} {\bibfnamefont {M.}~\bibnamefont {Neidig}}, \bibinfo {author}
  {\bibfnamefont {M.}~\bibnamefont {Ries}}, \bibinfo {author} {\bibfnamefont
  {A.}~\bibnamefont {Wenz}}, \bibinfo {author} {\bibfnamefont {G.}~\bibnamefont
  {Z{\"u}rn}}, \ and\ \bibinfo {author} {\bibfnamefont {S.}~\bibnamefont
  {Jochim}},\ }\href@noop {} {\bibfield  {journal} {\bibinfo  {journal}
  {Physical Review A}\ }\textbf {\bibinfo {volume} {90}},\ \bibinfo {pages}
  {043611} (\bibinfo {year} {2014})}\BibitemShut {NoStop}%
\bibitem [{Note1()}]{Note1}%
  \BibitemOpen
  \bibinfo {note} {Measurements performed at interaction strengths of up to
  ${\protect \rm ln}(k_{\protect \rm F}a_{\protect \rm 2D}) = 1.1$ show the
  same qualitative behaviour \cite {Supp}, but we currently do not have a
  quantitatively accurate temperature determination for our homogenous system
  at these interaction strengths.}\BibitemShut {Stop}%
\bibitem [{\citenamefont {Bohlen}\ \emph {et~al.}(2020)\citenamefont {Bohlen},
  \citenamefont {Sobirey}, \citenamefont {Luick}, \citenamefont {Biss},
  \citenamefont {Enss}, \citenamefont {Lompe},\ and\ \citenamefont
  {Moritz}}]{bohlen2020sound}%
  \BibitemOpen
  \bibfield  {author} {\bibinfo {author} {\bibfnamefont {M.}~\bibnamefont
  {Bohlen}}, \bibinfo {author} {\bibfnamefont {L.}~\bibnamefont {Sobirey}},
  \bibinfo {author} {\bibfnamefont {N.}~\bibnamefont {Luick}}, \bibinfo
  {author} {\bibfnamefont {H.}~\bibnamefont {Biss}}, \bibinfo {author}
  {\bibfnamefont {T.}~\bibnamefont {Enss}}, \bibinfo {author} {\bibfnamefont
  {T.}~\bibnamefont {Lompe}}, \ and\ \bibinfo {author} {\bibfnamefont
  {H.}~\bibnamefont {Moritz}},\ }\href@noop {} {\bibfield  {journal} {\bibinfo
  {journal} {arXiv preprint arXiv:2003.02713}\ } (\bibinfo {year}
  {2020})}\BibitemShut {NoStop}%
\bibitem [{\citenamefont {Hueck}\ \emph {et~al.}(2017)\citenamefont {Hueck},
  \citenamefont {Luick}, \citenamefont {Sobirey}, \citenamefont {Siegl},
  \citenamefont {Lompe}, \citenamefont {Moritz}, \citenamefont {Clark},\ and\
  \citenamefont {Chin}}]{hueck2017calibrating}%
  \BibitemOpen
  \bibfield  {author} {\bibinfo {author} {\bibfnamefont {K.}~\bibnamefont
  {Hueck}}, \bibinfo {author} {\bibfnamefont {N.}~\bibnamefont {Luick}},
  \bibinfo {author} {\bibfnamefont {L.}~\bibnamefont {Sobirey}}, \bibinfo
  {author} {\bibfnamefont {J.}~\bibnamefont {Siegl}}, \bibinfo {author}
  {\bibfnamefont {T.}~\bibnamefont {Lompe}}, \bibinfo {author} {\bibfnamefont
  {H.}~\bibnamefont {Moritz}}, \bibinfo {author} {\bibfnamefont {L.~W.}\
  \bibnamefont {Clark}}, \ and\ \bibinfo {author} {\bibfnamefont
  {C.}~\bibnamefont {Chin}},\ }\href@noop {} {\bibfield  {journal} {\bibinfo
  {journal} {Optics express}\ }\textbf {\bibinfo {volume} {25}},\ \bibinfo
  {pages} {8670} (\bibinfo {year} {2017})}\BibitemShut {NoStop}%
\bibitem [{\citenamefont {Z{\"u}rn}\ \emph {et~al.}(2013)\citenamefont
  {Z{\"u}rn}, \citenamefont {Lompe}, \citenamefont {Wenz}, \citenamefont
  {Jochim}, \citenamefont {Julienne},\ and\ \citenamefont
  {Hutson}}]{zurn2013precise}%
  \BibitemOpen
  \bibfield  {author} {\bibinfo {author} {\bibfnamefont {G.}~\bibnamefont
  {Z{\"u}rn}}, \bibinfo {author} {\bibfnamefont {T.}~\bibnamefont {Lompe}},
  \bibinfo {author} {\bibfnamefont {A.~N.}\ \bibnamefont {Wenz}}, \bibinfo
  {author} {\bibfnamefont {S.}~\bibnamefont {Jochim}}, \bibinfo {author}
  {\bibfnamefont {P.}~\bibnamefont {Julienne}}, \ and\ \bibinfo {author}
  {\bibfnamefont {J.}~\bibnamefont {Hutson}},\ }\href@noop {} {\bibfield
  {journal} {\bibinfo  {journal} {Physical review letters}\ }\textbf {\bibinfo
  {volume} {110}},\ \bibinfo {pages} {135301} (\bibinfo {year}
  {2013})}\BibitemShut {NoStop}%
\bibitem [{\citenamefont {Shi}\ \emph {et~al.}(2015)\citenamefont {Shi},
  \citenamefont {Chiesa},\ and\ \citenamefont {Zhang}}]{shi2015ground}%
  \BibitemOpen
  \bibfield  {author} {\bibinfo {author} {\bibfnamefont {H.}~\bibnamefont
  {Shi}}, \bibinfo {author} {\bibfnamefont {S.}~\bibnamefont {Chiesa}}, \ and\
  \bibinfo {author} {\bibfnamefont {S.}~\bibnamefont {Zhang}},\ }\href@noop {}
  {\bibfield  {journal} {\bibinfo  {journal} {Physical Review A}\ }\textbf
  {\bibinfo {volume} {92}},\ \bibinfo {pages} {033603} (\bibinfo {year}
  {2015})}\BibitemShut {NoStop}%
\end{thebibliography}%

\clearpage
\pagebreak
\begin{center}
\textbf{\large Supplementary materials}
\end{center}
\setcounter{figure}{0}
\renewcommand{\theequation}{S\arabic{equation}}
\renewcommand{\thefigure}{S\arabic{figure}}
\renewcommand{\thetable}{S\arabic{table}}

\subsection*{Preparation Scheme and Tuning of Interactions}
\begin{figure}[htbp]
\begin{center}
\includegraphics[width=\linewidth]{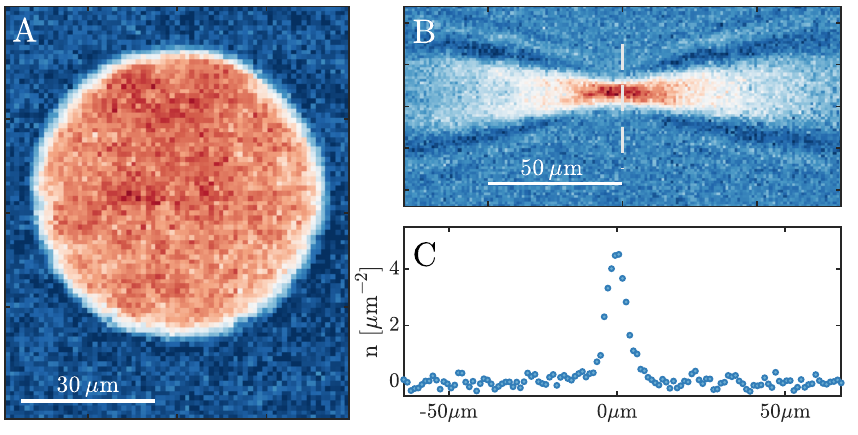}
\caption{\textbf{Imaging a homogeneous 2D Fermi gas}
(A) In-situ absorption image of the density distribution in our box potential \cite{hueck2017calibrating}. 
(B) Momentum distribution after performing matter wave focusing by letting the gas expand into a harmonic potential for a quarter of the trap period.
The tight confinement in the z-axis leads to a rapid expansion of the gas in this direction, which causes the gas to expand far outside the depth of field of the imaging system.
By tilting the imaging beam relative to the z-axis, the image of the gas at different z-positions is displaced laterally, with the gas being in focus in the center and imaging aberrations increasing to the sides.
(C) In-focus part of the momentum distribution along the dashed grey line in (B). 
The condensate peak at low momenta is clearly visible.
All images are the result of averaging over 8 individual measurements.
Note that resolving thermal wings in the momentum distribution (C) requires averaging over a significantly higher number of images.
}
\label{SM_Fig_Images}
\end{center}
\end{figure}

We perform our experiments with a balanced spin mixture of $^6{\rm Li}$ atoms in the $\left|F=1/2, m_F =1/2\right>$ and $\left|F=1/2, m_F =-1/2\right>$ hyperfine states.
The atoms are trapped in a box potential, resulting in a homogeneous density distribution as shown in Fig. \ref{SM_Fig_Images}.
As the critical velocity depends on the density of the gas, this homogenous density is critical for observing a sharp onset of dissipation at $v_{\rm c}$. 
 
The experimental setup and the procedure used to prepare homogeneous 2D Fermi gases are described in detail in \cite{hueck2018two}. 
In brief, we first prepare an ultracold gas of $^6{\rm Li}$ atoms in a highly elliptical optical dipole trap. 
We then perform further evaporative cooling and transfer the remaining $4000-8000$ atoms into a circular box potential with a diameter of $D \approx 60\,\rm{\mu m}$.

To bring the system into the 2D regime, the atoms are confined in the z-direction in a single antinode of an optical standing wave potential.
This potential is created by two blue-detuned ($\lambda=532\,\rm{nm}$) laser beams interfering under a shallow angle, resulting in an optical lattice with a lattice spacing of approximately $3\,\rm{\mu m}$ and a harmonic oscillator spacing of $\hbar \omega_z \approx h \,9.2\,\rm{kHz}$.
For all measurements shown in this work, both the chemical potential $\mu$ and the thermal energy $k_{\rm B} T$ of the gas were kept well below the level spacing $\hbar \omega_z$, thus avoiding population of excited states in the z-direction.
Therefore, we can parametrize the interparticle interactions by an effective 2D scattering length $a_{\rm 2D}$ and treat the gas as an effective 2D system.
The 2D scattering length depends on the harmonic oscillator length in z-direction $l_{\rm z} = \sqrt{\hbar/m \omega_{\rm z}}$ and the 3D scattering length $a_{\rm 3D}$ according to $a_{\rm 2D}=l_{\rm z}\sqrt{\pi/0.905} \exp(-\sqrt{\pi/2} \cdot l_{\rm z}/a_{\rm 3D})\exp(-\frac{1}{2}\Delta w)$ \cite{petrov2001interatomic}, where $\Delta w(\mu/\hbar\omega_z)$ is a momentum-scale correction that becomes relevant on the BCS side of the crossover.
In our experiments we tune the 2D scattering length $a_{\rm 2D}$ by varying $a_{\rm 3D}$ using a broad Feshbach resonance located at a magnetic field of $B =832$\,G \cite{zurn2013precise}. 
This allows us to continuously tune the system from a gas of deeply bound dimers to a BCS superfluid.

\begin{table}
\begin{center}
\begin{tabular}{ l | r | r | r | r }
     & $\mu/h$ [kHz] & $n$ [$\mu\rm{m}^{-2}$]  & $V_{\rm {Latt}}/h$ [kHz] \\
    \hline     
    Fig. \ref{Fig2}A & 0.96  & 1.5  & 0.36\\ 
    Fig. \ref{Fig2}B & 5.1 & 0.80  & 0.58 \\ 
    Fig. \ref{Fig3} & 0.66 & 1.1  & 0.24 \\ 
    Fig. \ref{Fig4} & - & 0.80 & 0.49
\end{tabular}
\end{center}
\caption{\textbf{Experimental parameters.}
		Chemical potential $\mu$, density $n$ and lattice power $V_{\rm {Latt}}$ used to obtain the data shown in Figs. \ref{Fig2}-\ref{Fig4}. 
    The value of the chemical potential was calculated from the measured density using the equation of state published in \cite{shi2015ground}.
   	The height of the lattice potential is determined from the laser power of the lattice beams using the mean of the two calibration results described in the text.}
    \label{SM_Tab_Parameters}
  \end{table}

\subsection*{Lattice Calibration}

To observe frictionless flow in our 2D Fermi gas, we realize Landau's gedankenexperiment of a mobile impurity moving through the system without dissipation.
In Landau's scenario, this disturbance is point-like and can excite the system at all momenta.
However, when trying to experimentally realize this with a focused laser beam, the shape and finite size of the focus introduce a momentum scale that is difficult to control.
Hence, we use a moving optical lattice as our impurity, since it has a well-defined and tunable momentum transfer that is determined by the lattice wavevector $k_0$.

The moving optical lattice is created by interfering two red-detuned ($\lambda = 780\,\rm{nm}$) laser beams with a controllable frequency difference. 
To obtain these beams, we use light from an extended-cavity diode laser (Toptica DL PRO 780), split it into two paths and route each beam through an independently controlled acousto-optical modulator (AOM). 
This allows us to create an optical lattice moving at a speed of $v=L \Delta\nu$ by setting the frequency difference $\Delta\nu$ between the two AOMs.
We can also tune the momentum transfer $\hbar k_0 \propto 1/L $ of the lattice by varying the distance of the two beams on the entrance aperture of the high-resolution objective that focuses them onto the atoms.
This changes the crossing angle $\alpha$ of the interfering beams and thus the lattice spacing $L = \frac{\lambda}{2 \sin{\alpha/2}}$. 

To determine the lattice spacing, we image the potential directly onto a camera using a second high-resolution objective. 
The height of the lattice potential is calibrated by projecting each beam onto the gas individually and measuring the change in the density distribution as a function of laser power.
We observe a linear dependence of the change in density on the laser power and use the known equation of state of the system to extract the potential height as a function of laser power for both beams. 
We find potential heights $V/h$ of $108\,\rm{Hz/mW}$ and $122\,\rm{Hz/mW}$ for the two beams.
As the contrast of the interference pattern can be extracted from the images of the intensity distribution and is $\gtrsim 0.95$, the potential height is computed to be $V_{\rm {Latt}}/h \approx 460\,\rm{Hz/mW}$.

Alternatively, we can project the optical lattice at the widest lattice spacing onto the atoms at variable laser power and directly measure the amplitude of the resulting density modulation. 
Using this method, we obtain $V_{\rm {Latt}}/h \approx 510\,\rm{Hz/mW}$, showing reasonable agreement between the two methods. 
Thus, the lattice heights used in this work are a small fraction of the chemical potential of the gas for all but the lowest values of $\rm{ln}(k_{\rm F}a_{\rm 2D})$ (see Table \ref{SM_Tab_Parameters}).

\subsection*{Thermometry and Controlled Heating}

Performing thermometry on strongly interacting degenerate 2D Fermi gases is challenging, as there is only very limited theory available for these systems.
For harmonically trapped gases, this problem can be circumvented since a significant fraction of the atoms is in the low density region at the edge of the trap where the gas is non-degenerate and can be reasonably well described by a Boltzmann distribution.
This allows the extraction of the temperature of the gas from the \textit{in situ} density distribution as done for example in \cite{boettcher2016equation}.
For our homogeneous system, however, such low density wings do not exist.

An alternative approach is to use matter wave focusing, where a weak harmonic confinement is used to perform a rotation in phase space \cite{murthy2014matter, hueck2018two} to extract the momentum distribution of the system. 
In our case, the harmonic confinement has a trap frequency of $\omega_{\rm mag} \approx 2\pi\cdot 28\,\rm{Hz}$ and is provided by the curvature of the magnetic offset field.
Since this technique requires ballistic expansion of the sample, we can only use it in the BEC regime, where the interactions are weak enough that the effect of collisions during the time of flight can be neglected.

As shown in \cite{boettcher2016equation}, the high-momentum tail of the momentum distribution is well described by a Boltzmann equation of state $n\lambda_T^2 = e^{\mu_d/kT}$, where $\lambda_T = \sqrt{\frac{2\pi\hbar^2}{m_dk_{\rm B}T}}$ is the thermal wavelength of dimers with mass $m_d$ and $\mu_d$ is the chemical potential of the dimers. 
Hence we can extract the temperature of a gas in the BEC regime by performing matter wave focusing and fitting the high-momentum part of the momentum distribution. 
However, since the signal-to-noise ratio for this high-momentum part is quite low in our measurements, this method requires considerable averaging and is not suitable as a single-shot thermometer. 
We therefore use the change of the height of the condensate peak, which can be determined with a much higher signal-to-noise ratio, to quantify the response of the system to the moving periodic potential. 

\begin{figure}[htbp]
\begin{center}
\includegraphics[width=\linewidth]{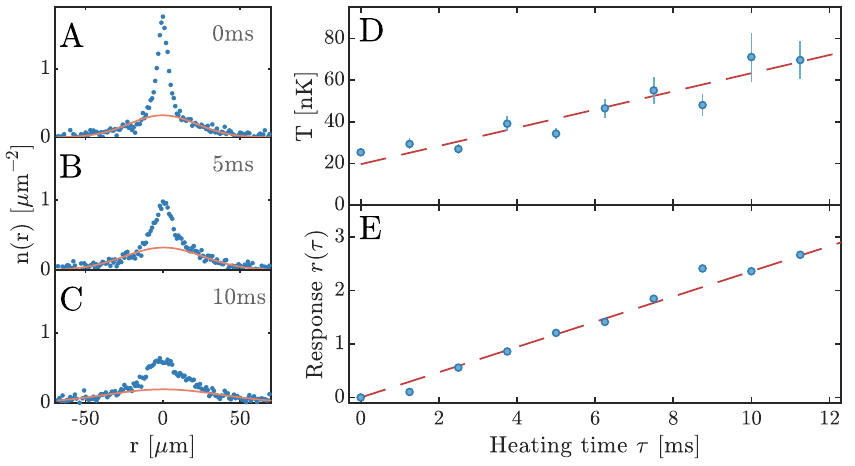}
\caption{\textbf{Calibration of the heating procedure.}
(A-C) Measurements of the momentum distribution of the gas after applying the moving lattice to the system for different heating times $\tau$.
The red solid lines show the Boltzmann fits to the wings of the distribution from which the temperature is extracted.
(D) Temperature $T$ of the gas as a function of heating time. 
We observe that the temperature of the gas increases roughly linearly with the heating time (red dashed line). 
(E) Response $r(\tau)$ as a function of heating time. 
The response shows a very similar behavior to the temperature, but can be measured with a much higher signal-to-noise ratio. 
All data points are obtained from an average of $55$ individual measurements.
The error bars in (D) denote the 1$\sigma$ confidence interval of the fit.}
\label{SM_Fig_Therm}
\end{center}
\end{figure}

To increase the temperature of the gas in a controlled manner, we move the periodic potential through the system for a variable heating time $\tau$ at a velocity larger than the critical velocity. 
The resulting change in the momentum distribution of the system is shown in Fig. \ref{SM_Fig_Therm}\,A-C, with the extracted temperatures shown in Fig. \ref{SM_Fig_Therm}\,D.
For comparison, we show the change in the height of the condensate peak, plotted as $r(\tau) = (n(k=0,\tau=0)/n(k=0,\tau))-1$ in Fig. \ref{SM_Fig_Therm}\,E.
We use this measurement to calibrate the heating procedure and thereby the temperature axis shown in Fig. \ref{Fig3}\,D.
 
To obtain an estimate of the systematic uncertainty of our determination of  $T_{\rm c}$, we performed a second set of measurements with a homogeneous 2D Fermi gas in a different potential: Using a red-detuned ($\lambda = 770\,\rm{nm}$) flat-bottom potential shaped by a digital micromirror device (DMD) instead of blue-detuned walls creates a box trap of the same dimensions as in the previous measurement, but with a weak ($\omega_{\rm mag} \approx 2\pi\cdot 28\,\rm{Hz}$) harmonic confinement outside the trap volume that is populated by a small number of thermal atoms. 
While the lowest temperatures we achieved in this red-detuned potential were significantly higher than in the blue-detuned trap, we were able to perform a measurement of the critical temperature at comparable interaction strength and density to the data shown in Fig. \ref{Fig3} of the main text. 
For this measurement, thermometry was performed using a Boltzmann fit to the low density part of the \textit{in situ} density distribution, from which we obtain a critical temperature of $T_{\rm c} = 0.116(2)\,T_{\rm F}$ using the same evaluation as for the data shown in Fig. \ref{Fig3}. 
A source of error on the temperatures extracted using these \textit{in situ} fits are inhomogeneities of the harmonic potential, for example due to the optical lattice used for the 2D confinement, as well as small offsets in the absorption imaging, which most likely lead to an overestimation of the temperature of the gas.
As a rough approximation for the systematic error of our determination of $T_{\rm c}$ we use the difference between the values obtained using time-of-flight thermometry ($T_{\rm c}/T_{\rm F} = 0.094(4) $) and \textit{in situ} thermometry ($T_{\rm c}/T_{\rm F} = 0.116(2)$).
This then yields a critical temperature of $T_{\rm c}/T_{\rm F} = 0.094 \pm 0.004_{\rm stat} \pm 0.02_{\rm sys}$ for a bosonic system at an interaction parameter of $\rm{ln}(k_{\rm F}a_{\rm 2D}) = -2.9$.

\subsection*{Critical Temperature in the BEC-BCS Crossover}

\begin{figure}[hbp]
\begin{center}
\includegraphics[width=\linewidth]{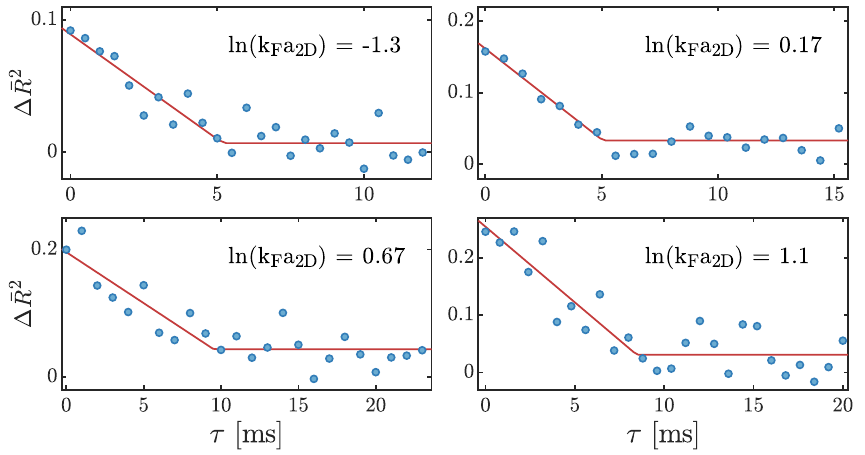}
\caption{\textbf{Superfluid phase transition at different interaction strengths.}
Difference $\Delta \bar{R}^2$ of the adjusted coefficients of determination between fits to $r(v,T)$ with zero and non-zero $v_{\rm c}$ as a function of heating time $\tau$ for different interaction strengths. 
Linear fits of $\Delta \bar{R}^2$ with a threshold at a critical heating time (red lines) clearly show the appearance of a non-zero critical velocity below a critical temperature for systems across the crossover (A,B,C) and into the BCS regime (D).  
However, due to the challenges involved in measuring the temperature of strongly interacting homogenous Fermi gases discussed in the text we are currently not able to quantitatively determine the relation between the heating time and the temperature for these systems and hence cannot give a value for $T_{\rm c}$.
}
\label{FigSMTc}
\end{center}
\end{figure}

To observe the phase transition from a superfluid to a normal state across the BEC-BCS crossover, we performed measurements similar to the one shown in Fig. \ref{Fig3} for interaction strengths ranging from $\rm{ln}(k_{\rm F}a_{\rm 2D}) = -4.1$ deep in the molecular regime to $\rm{ln}(k_{\rm F}a_{\rm 2D}) = 1.1$ on the BCS side of the resonance (Fig. \ref{FigSMTc}).
For these measurements, we prepare and heat the system at the same interaction strength of $\rm{ln}(k_{\rm F}a_{\rm 2D}) = -2.9$ as for the measurements shown in Fig. \ref{Fig3}, ramp to different magnetic fields to tune the interaction strength to the desired value, and perform measurements of $v_{\rm c}$ at these interaction strengths. 
We observe a clear qualitative difference in the response to the moving lattice between cold and hot systems at all measured interaction strengths, showing the presence of a critical temperature.
However, as the temperature of the gas changes during the interaction ramp and we cannot use matter wave focusing to determine the temperature at higher interaction strengths, we are currently unable to quantitatively determine the critical temperatures of these systems.
Nevertheless, these results present a promising starting point for a future measurement of the critical temperature for superfluidity in 2D Fermi gases across the BEC-BCS crossover.

\end{document}